\documentstyle[aps,twocolumn]{revtex}

\newcommand {\be}{\begin{equation}}
\newcommand {\ee}{\end{equation}}

\begin{document}

\title{Towards a Statistical Theory
of Finite Fermi Systems and Compound States:\\ 
Random Two-Body Interaction Approach}

\author{V.V.Flambaum$^{1,2}$,
F.M.Izrailev$^{2,3}$\thanks{email addresses: izrailev@vxinpb.inp.nsk.su; 
izrailev@physics.spa.umn.edu}, 
and G.Casati$^3$}

\address{ $^1$ School of Physics, University of New South Wales,
Sydney, 2052, Australia}

\address{ $^2$ Budker Institute of Nuclear Physics,
630090 Novosibirsk, Russia}

\address{ $^3$ Universita di Milano, sede di Como, via Lucini
3, 22100 Como, Italy\\
and Istituto Nazionale Fisica della Materia}

\twocolumn[ 
\date{\today}
 
\maketitle
 
\widetext
 
\vspace*{-1.0truecm}
 
\begin{abstract}
\begin{center}
\parbox{14cm}

The model of Fermi particles with random two-body interaction 
is investigated. This model allows to study
the origin and accuracy of statistical laws in
few-body systems, the role of interaction and chaos in
thermalization, Fermi-Dirac distribution for quasi-particles
with spreading widths, matrix elements of external field and
enhancement of weak perturbation in chaotic compound states.

\end{center}
\end{abstract}

\pacs{PACS numbers: 31.20.Tz, 31.50.+w, 32.30.-r, 05.45.+b }
] \narrowtext

As is known, excited states in many-body systems such as compound nuclei,
rare-earth atoms, molecules, atomic clusters, quantum dots in solids, etc.
are very complicated (''chaotic'') and can be described via
statistical methods. There are two major approaches. The first is based on
random matrix theory (see e.g.
\cite{P65}). This approach is very general in nature and therefore it does
not describe many important properties of realistic many body systems, its
prediction being limited to level statistics, localization properties of
eigenstates and the like. The other approach is based on direct numerical
investigations of the given particular system. For example, the results of
direct diagonalization of the Hamiltonian matrix for the rare-earth Ce-atom
have been compared with statistical theory of compound states \cite{FGGK94}.
A similar study has been performed for the $s-d$ shell model of complex
nucleus \cite{ZELE} where the problem of thermalization has also been
considered. Quite obviously with this second approach, it is very hard (e.g.
due to the lack of statistics) to draw general conclusions concerning the
accuracy of statistical laws in systems with finite number of particles and
the conditions for their applicability.

Here we suggest an ''intermediate'' approach based on a simple mathematical
model with random interaction, which however takes into account the most
important features of many-body systems such as single-particle orbitals,
two-body interaction and the Pauli principle. In our model of Random
Two-Body Interaction (RTBI) we assume that the system consists of $n$
Fermi-particles which can occupy $m\,$ orbitals with no double occupancy
(extension to Bose-systems is straightforward). As an example, the energies
of orbitals are taken in the form 
\begin{equation}
\label{eps}\epsilon _\alpha=d(\alpha+\frac
1\alpha),\;\;\;\;\;\;\;\;\alpha=1,2,...,m 
\end{equation}
where the second term is introduced in order to avoid the degeneracy of
many-particle states. Matrix elements of two-body interaction are chosen as
random variables distributed according to the Gaussian law with the zero
mean and the variance $V_0$ . Therefore, our model has $3$ independent
parameters: $n,m$ and $V_0/d$ .

{\bf Structure of the Hamiltonian matrix. }The basis of our Hamiltonian
matrix is chosen to correspond to the products of single-particle orbitals
ordered by increasing many-body unperturbed energies. Therefore, without
interaction the matrix is diagonal, with increasing elements. {\rm The size }%
$N$ {\rm of RTBI-matrix is given by }$N=C_n^m=m!/(n!(m-n)!)$ and is
exponentially large {\rm \ for large }$n$ and $m$ . Due to the two-body
character of interaction the matrix elements are zero between those basis
states which differ by more than two orbitals. Therefore, the Hamiltonian
matrix is essentially sparse with the sparsity (ratio of non-zero elements
to the total number $N^2$) given by $s=(1+n(m-n)+n(n-1)(m-n)(m-n-1)/4)/N$ .
At large $n,m$ the sparsity is exponentially small, $s\sim \exp (-n)$ . In
fact, the number $N_2=m^2(m-1)^2/4$ of independent variables (number of pair
interactions) in the model is even smaller than the number of non-zero
Hamiltonian matrix elements.

There are three types of interactions between many-body states: diagonal
interaction which contains $k_{int}=n(n-1)/2$ two-body terms, interaction
between states which differ by one orbital with $k_{int}=n-1$ terms and
interaction between states which differ by two orbitals with $k_{int}=1$ .
Finally, distant many-body states which differ by more than two orbitals
have zero matrix elements. Thus, the Hamiltonian matrix $H_{ij}$ has large
and increasing diagonal, plus sparse band-like structure with a decrease of
off-diagonal elements as a function of the distance $|i-j|\,$ from the
diagonal.

In what follows we consider a particular case which can be used to describe
Ce-atom : $m=11,\,n=4$ (therefore, $N=330$), $d=1,V_0=0.12$ (the two latter
parameters are given in eV). Direct investigation of this atom (see details
in \cite{FGGK94}) has shown that it can be treated as a very chaotic system.

{\bf Energy spectrum and eigenstates. } As was found in \cite{FW70} (see
also \cite{brody}), for two-body random interaction the density of states
(DOS) should be of Gaussian type, provided $m\gg n \gg 1$. Our numerical
data show that in the RTBI model in spite of small number of particles
the DOS is quite close to a Gaussian with 
$\left\langle E\right\rangle =25.1$ and $\sigma=5.7$.
Another commonly discussed characteristic of the spectrum
is the level spacing distribution; in our model it is well described by the
famous Wigner-Dyson law. This fact is also in agreement with old studies 
\cite{BF71} of some two-body interaction models (see also references in \cite
{brody}). However, our interest goes beyond the DOS and spectrum statistics.

A quantity of relevant interest is the so-called spectral local density of
states or strength function. It is defined by the weight $w(E,E_j)=\overline{%
|C_j^2|}\,$ of a particular basis component $j=1,2,..,N$ in the eigenstates
with an energy close to $E$ , according to the relation 
\begin{equation}
\label{ldos}\rho (j,E)=\frac{w(E,E_j)}D=
<\sum_r|C_j(E_r)|^2\delta (E-E_r)> 
\end{equation}
\ Here $D$ is the local mean level spacing. The width of this function in
energy $E$ is proportional to an effective number of components in the
expansion of an unperturbed basis state in terms of exact eigenstates. In
nuclear physics, this function is usually assumed to have the Breit-Wigner
form, $\rho (j,E)\sim \left[ (E-E_j)^2+\Gamma _c^2/4\right] ^{-1}$ (the
relevance of this function to ergodicity and chaos is studied in \cite{boris}%
). According to our numerical results, in the center of the spectrum $%
(j\approx N/2)$ the spreading width$\,$is equal to $\Gamma _c\approx 1.0.$
One should note that close to the edges $j=1,N$ , the symmetric form of the
distribution $\rho (j,E)$ is strongly distorted. However, the width of the
distribution itself changes slightly. What is more important, the tails of
the function $\rho (j,E)\,$ decay much faster than in the
Breit-Wigner law. This is the consequence of the band-like structure of the
Hamiltonian matrix. Both the value of $\Gamma _c$ and the fast decay of the
tails, are in agreement with Ce-atom calculations \cite{FGGK94}.

The localization length (number of principal components of the eigenstates)
can be defined through the relation $l\sim \exp (\left\langle S\right\rangle
)_{}$where $S$ is the statistical entropy of individual eigenstates (see,
e.g.\cite{I90}). One should note that in spite of a completely random
character of the interaction, the lowest states turn out to be quite simple
containing $l\approx 1$ basis components. This can be explained
by the low density of states near the ground state. The localization length $%
l$ is maximal in the center $E\approx \left\langle E\right\rangle $ of the
energy band and, according to our numerical data, can be well described by
the Gaussian function: $l=A\exp (-(E-\left\langle E\right\rangle
)^2/(2\sigma ^2))$ with $A\approx 135\,$ and $\sigma \approx 5.45$ .Thus, in
the center of the spectrum the number of principle components is about $100$
which is again very close to the results of direct calculations for Ce-atom 
\cite{FGGK94}.

{\bf Statistical} {\bf treatment of finite Fermi-systems. } As is known,
quantum statistical laws are derived for systems with infinite number of
degrees of freedom, or for systems in a thermostat. From this point of view,
it is of importance to study how statistical laws appear in systems with
finite number of particles. Below, we show that in the RTBI model one can
introduce a reasonably accurate statistical description based on such
macroscopic characteristics as the temperature $T$, the chemical potential $%
\mu $ and average occupations numbers $n_i\,$for the orbitals. In the mean
field approximation these quantities can be obtained from the following set
of equations: 
\begin{equation}
\label{n}\sum_{\alpha =1}^mn_\alpha =n;\,\,\,\,\,\sum_{\alpha =1}^m\epsilon
_\alpha n_\alpha +<\sum_{\alpha \geq \beta }^mV_{\alpha \beta }n_\alpha
n_\beta >\;=E 
\end{equation}
\begin{equation}
\label{occup}n_\alpha \equiv n(\epsilon _\alpha )=(1+\exp (\tilde \epsilon
_\alpha -\mu )/T)^{-1} 
\end{equation}
\begin{equation}
\label{quasi}\tilde \epsilon _\alpha =\epsilon _\alpha +<\sum_{\beta
=1}^mV_{\alpha \beta }\,n_\beta > 
\end{equation}
In our case, the average interaction $<V_{\alpha \beta }>$ is zero,
therefore, we can omit interaction term in (\ref{n}) and (\ref{quasi}). On
the other hand, there exist other important effects of the interaction which
appear beyond the mean field approximation. Namely, the interaction leads to
the spreading width for the basis states and for the single-particle
orbitals $(\Gamma _\alpha )$. It also results in the shift of average
energies, $\tilde \epsilon _\alpha =\epsilon _\alpha +\delta \epsilon
_\alpha $ . According to our numerical data, the magnitudes $\delta \epsilon
_\alpha $ are smaller than $\Gamma _\alpha $ and vanish in the mean ($\delta
\epsilon _\alpha <0$ for low orbitals, $\delta \epsilon _\alpha >0\,\,$ for
high orbitals and $\delta \epsilon _\alpha \approx 0$ near the center). For
this reason, we will take into account the effect of the spreading width $%
\Gamma _\alpha $ only.

Instead of (\ref{occup}), by averaging the Fermi-Dirac distribution $%
n(\epsilon )$ over the interval $\Gamma _\alpha $: 
\begin{equation}
\label{ngamma}n_\alpha =\int\limits_{\epsilon _\alpha -\Gamma _\alpha
/2}^{\epsilon _\alpha +\Gamma _\alpha /2}n(\epsilon )\frac{d\epsilon }{%
\Gamma _\alpha }=1-\frac T{\Gamma _\alpha }\ln \left[ \frac{1+\exp \frac{%
(\epsilon _\alpha +\frac{\Gamma _\alpha }2-\mu )}{2T}}{1+\exp \frac{%
(\epsilon _\alpha -\frac{\Gamma _\alpha }2-\mu )}{2T}}\right] 
\end{equation}
we now introduce the occupation numbers (\ref{ngamma}) which take into
account the finite spreading width of ''quasi-particles''. In the limit $%
\Gamma _\alpha =0$ the expression (\ref{occup}) with $n_\alpha =n(\epsilon
_\alpha )$ is recovered. The numerical solution of Eqs. (\ref{n},\ref{ngamma})
is presented in Fig.1. In order to reveal the influence of the spreading
width $\Gamma _\alpha $, two curves are given for comparison with $\Gamma
_\alpha =0$ and $\Gamma _\alpha =3.0$ . The latter value of $\Gamma _\alpha
\,$was deliberately taken larger than the spreading width of basic
components $\Gamma _c\approx 1.0$ in order to elucidate the effect of $%
\Gamma _\alpha $ . The temperature $T$ and chemical potential $\mu $ are
calculated from the above equations (\ref{n},\ref{ngamma}) at fixed
excitation energy $\delta E=E-E_{min}\,.$ We found that for small $\delta E$
the temperature is quite different for different values of ${\Gamma _\alpha }
$. Surprisingly, in spite of this, the two ''theoretical'' curves
practically coincide. This means that the temperature mimics the statistical
effect of the interaction, the phenomenon far from being trivial.

The average occupation numbers directly computed from exact eigenstates are
shown in Fig.1 with circles. One can see that even for $4$ particles the
actual distribution of occupation numbers can be approximately 
described by the statistical
Fermi-Dirac distribution. However, there is a clear deviation which  
indicates that for the chosen parameters 
the thermalization is not complete. Numerical data show that the deviation 
disappears
with increasing the excitation energy $\delta E$ , while it increases when the
perturbation $V_0\,$ decreases. The latter effect has been also observed in
the $s-d$ nuclear shell model \cite{ZELE} where the relevance of chaos to
the thermalization was studied in a different approach. We have to stress
that equilibrium distribution, or ''thermalization'', in this few--particle
system is due to the interaction which ''chaotically'' mixes neighboring
basis states with different occupation numbers. To demonstrate this, we plot
in Fig.1 the occupations numbers for the same system with no interaction $%
(V_0=0)$ . Their distribution has nothing to do with the Fermi-Dirac
distribution: it is singular and for some values of $\delta E$ even not
monotonous.

{\bf Correlations between occupation numbers. }In a few body-system one
could expect quite strong correlations between occupation numbers of
different orbitals. We have found, instead, that typically the correlations 
are weak; even for close orbitals the ratio $\left\langle
\hat n_\alpha \right\rangle \left\langle \hat n_\beta \right\rangle /
\left\langle \hat n_\alpha \hat n_\beta \right\rangle \approx
0.82$. Only when occupation numbers are small, the correlations 
can be very strong.

{\bf Matrix elements of an external perturbation. }The main problem in the
compound state theory is the calculation of effects of an external
perturbation. Since matrix elements of any single-particle operator can be
expressed in terms of elementary density matrix operators $\hat \rho
_{\alpha \beta }=a_\alpha ^{\dagger }a_\beta $ which transfer the particle
from the orbital $\beta $ to the orbital $\alpha $ , the main interest is in
statistical properties of matrix elements $\rho _{\alpha \beta }$ . Below,
we use recent approach developed in \cite{F} (see also the study of Ce-atom
in \cite{FGGK94}) where the following expression for the mean square matrix
elements (MSME) $\xi _{\alpha \beta }^{(n_2n_1)}\equiv \overline{\left| \rho
_{\alpha \beta }^{(n_2n_1)}\right| ^2}$ between compound states $%
\left\langle n_1\right| $and $\left\langle n_2\right| $has been derived: 
\begin{equation}
\label{rho}\xi _{\alpha \beta }^{(n_2n_1)}=Q_{\alpha \beta} 
\sum\limits_rw_{n_1}(E_r+%
\omega _{\alpha \beta })w_{n_2}(E_r) 
\end{equation}
where $\omega _{\alpha \beta }=\epsilon _\alpha -\epsilon _\beta $
and $Q_{\alpha \beta}= \overline{\left\langle n_2\left|
\hat n_\alpha (1-\hat n_\beta )\right|n_2\right\rangle }$. Here the
sum runs over many-body basis components with $w_{n_1},w_{n_2}$ being 
the weights of
these components in the states $n_1$ and 
$n_2$ (see Eq. (\ref{ldos})). If we know the shape of the strength
function $w$, this sum can be replaced by an integral and explicitly
evaluated. In particular, when $w$ iis Breit-Wigner type, the final
expression for the MSME has also the Breit-Wigner form (see details in \cite
{FGGK94}).

In order to check the accuracy of the above expression (\ref{rho}) in the
RTBI model, we have directly calculated $\xi _{\alpha \beta }^{(n_2n_1)}$
for $\alpha =4;\,\beta =5$ (transition from the ground state to the nearest
one) and different values of $n_1$ and $n_2$. 
To reduce fluctuations, an averaging over a number of
realizations of the Hamiltonian matrix has been done. The direct comparison
with the analytical prediction (\ref{rho}) shows a quite good agreement.
In particular, the
positions of maximum as well as the widths of $\xi _{\alpha \beta
}^{(n_2n_1)}$ in dependence on $n_1$ and 
$n_2$ are well described by (\ref{rho}). However, quite unexpected
deviations have been discovered which were found to be generic in the model. A
typical example is given in Fig.2 where a clear difference is seen between
statistical approximation (\ref{rho}) and numerical data. The absolute
difference is maximal at the center of the $\xi $-dependence; however, in
the tails the relative difference is even larger. The thoroughly study of
this phenomenon (see details in \cite{FGI95}) has revealed a very intriguing
fact: the origin of this effect is in the underlying correlations induced by
the two-body nature of interaction. Similar correlations were observed in
the model of random separable interaction \cite{F94,FG95}. Full analytical
treatment of the correlations for the RTBI model is given in \cite{FGI95}.
In particular, for the tails the contribution of the correlation term $\xi
_{corr}$ to the total MSME, $\xi _{total}=\xi _{corr}+\xi _{stat}$, has been
estimated as 
\begin{equation}
\label{estimate}R\equiv \frac{\xi _{corr}}{\xi _{stat}}=-\frac{%
(n-2)(m-n-1)(m-n+2)}{n(m-n)(m-n+3)} 
\end{equation}
where $\xi _{stat}$ is very close to that given by the expression (\ref{rho}%
). In the maximum of $\xi $, the estimate for $m-n \gg 1$ 
reads as $\xi _{total}/\xi
_{stat}\approx 2-2m/(n(m-n))$. The remarkable result is that these
correlations do not decrease with an increase of number of particles $n$ and
number of orbitals $m$. Numerical data for larger values of $n=7$ and $m=14$
(with $N=3432$) have confirmed this prediction.

{\bf Enhancement of a weak perturbation in chaotic many-body systems. } The
RTBI model allows for the study of a very important effect, namely, the
enhancement of a weak perturbation $\hat W$, which takes place in systems
with chaotic compound states. This effect is proportional to the mixing
coefficient $\eta =<n_1|\hat W|n_2>/\Delta _{12}$ where $\Delta
_{12}=E_1-E_2 $ is the spacing between the neighboring energy levels of
compound states. Since the spacings for strongly excited states of many-body
systems are exponentially small, $\Delta _{12}\sim \exp (-n)$ , one could
expect strong enhancement of the perturbation in comparison with the
single-particle mixing defined by $\eta _s=<\alpha |\hat W|\beta >/(\epsilon
_\alpha -\epsilon _\beta )$. The possibility of the enhancement in compound
nuclei was pointed out in \cite{7,SF82} and considered in details in recent
review \cite{FG95}. However, such non-trivial effects like repulsion between
energy levels and the above discussed correlations, may have strong
influence on the enhancement. Our preliminary results show that even for $4$
particles the enhancement in the RTBI model does exist.

{\bf Conclusive remarks}. In this Letter we have analysed the RTBI model
which, unlike conventional random matrix models, allows for the
study of many
important problems related to the two-body character of the interaction. As an
example, the role of interaction for the appearance of the Fermi-Dirac
distribution has been investigated for the parameters of Ce atom. It was
found that two-body interaction gives rise to thermalization and that the
statistical effect of interaction can be imitated by an increase of
temperature. The study of correlations between occupation numbers has
revealed sufficiently weak correlations even for small number $(n=4)$ of
particles. This justifies the approximation of independent particles which
is typically used in the description of complex compound states. To describe
the effect of spreading widths of orbitals, the generalization of the
Fermi-Dirac distribution was suggested.

The numerical and analytical treatment has shown that the statistical theory
reproduces quite well the global structure of matrix elements of an external
perturbation between compound states. On the other hand, underlying
correlations have been discovered which are induced by the two-body
character of interaction, even if the latter is completely random. This
phenomenon results in serious deviations from the statistical predictions for
matrix elements between compound states. At the moment, all consequences of
these correlations are still not understood, however, preliminary data \cite
{FGI96} show that they can lead to the so-called gross-structure (sharp
peaks) in a cross section.

In conclusion, our results show that the discussed above TBRI model can be 
a very useful tool in the study of many important problems of statistical 
physics of complex quantum systems.

{\bf Acknowledgments} We are thankful to V.Sokolov and V.Zelevinsky for the
discussion and valuable comments. F.M.I. wishes to acknowledge the supports
of Grant ERBCHRXCT 930331 Human Capital and Mobility Network of the European
Community and also Grant No RB7000 from the International Science Foundation.




\vspace{1.0cm}

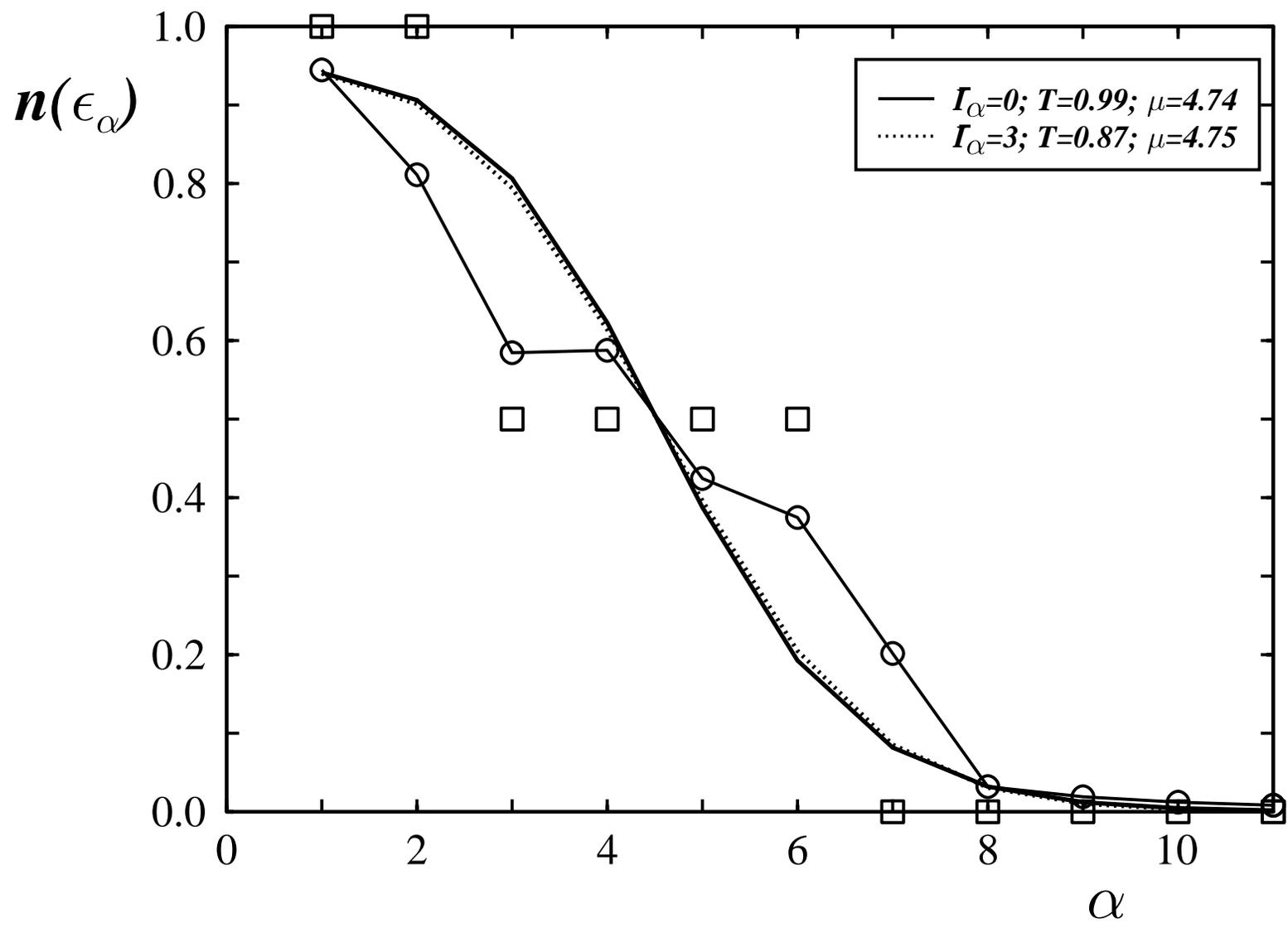
\noindent FIG.1. Distribution of occupation numbers for the excitation
energy $\delta E = 1.63$ averaged over an ensemble of 5 Hamiltonian 
matrices. The full and dotted lines are numerical solutions
of Eqs. (\ref{n},\ref{ngamma}) with $\Gamma_\alpha =0$ and $\Gamma_\alpha =
3 $ respectively. The circles and squares give results of the direct
computation of average occupation numbers with and without interaction
respectively. \\

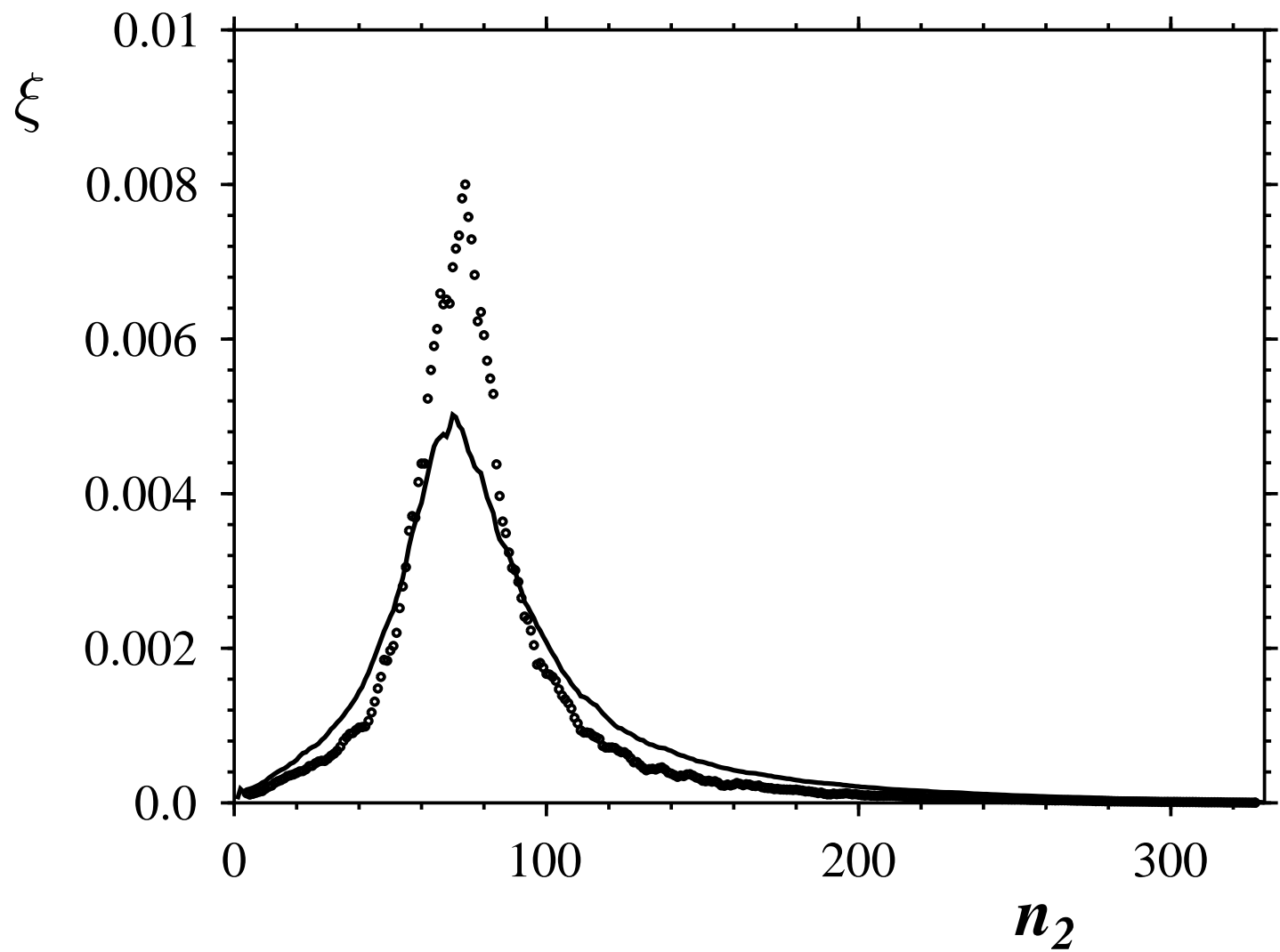
\noindent FIG.2. Mean square matrix element $\xi $ calculated for $%
n_1=55,\alpha =4,\beta =5$ as a function of $%
n_2 $. Averaging over $100$ Hamiltonian matrices with
different realization of random two-body interaction has been made. The dots
correspond to the direct numerical computation, the solid line represents
the statistical approximation (\ref{rho}).

\end{document}